\documentclass[pdflatex,sn-nature,iicol]{sn-jnl}

\usepackage{graphicx}%
\usepackage{multirow}%
\usepackage{amsmath,amssymb,amsfonts}%
\usepackage{amsthm}%
\usepackage{mathrsfs}%
\usepackage[title]{appendix}%
\usepackage{xcolor}%
\usepackage{textcomp}%
\usepackage{manyfoot}%
\usepackage{booktabs}%
\usepackage{algorithm}%
\usepackage{algorithmicx}%
\usepackage{algpseudocode}%
\usepackage{listings}%
\usepackage{comment}
\usepackage{siunitx}
\usepackage{placeins}

\theoremstyle{thmstyleone}%
\theoremstyle{thmstyletwo}%

\theoremstyle{thmstylethree}%

\newcommand{\UCRY}{\mathrm{UCR}_y}
\newcommand{\UCRZ}{\mathrm{UCR}_z}
\newcommand{\ee}{\mathrm{e}}

\usepackage{comment}
\usepackage{braket}
\begin{document}

\title{Optimal Quantum-Circuit Synthesis and Implementation of General Unitaries with Parallelized Bitstring-Conditioned Rotations}

\author*[1]{\fnm{Heikki} \sur{Suominen}}\email{heikki.suominen@aalto.fi}

\author[1,2]{\fnm{Suman} \sur{Kundu}}

\author[1]{\fnm{Heidi} \sur{Kivijärvi}}
\author[1,3]{\fnm{Sasu} \sur{Tuohino}}
\author[1]{\fnm{Mikko} \sur{Tuokkola}}
\author[1]{\fnm{Eemeli} \sur{Forsbom}}
\author[4]{\fnm{Madhavi} \sur{Chand}}%
\author[1,2]{\fnm{Vasilii} \sur{Vadimov}}%
\author[4]{\fnm{R.} \sur{Vijay}}%
\author[1,5]{\fnm{Mikko} \sur{Möttönen}}%


\affil[1]{\orgdiv{QCD Labs, QTF Centre of Excellence, Department of Applied Physics}, \orgname{Aalto University}, \orgaddress{\street{Tietotie 3}, \city{Espoo}, \postcode{02150}, \country{Finland}}}
\affil[2]{\orgname{Coherence Quantum}, \orgaddress{\street{Metsänneidonkuja 10}, \city{Espoo}, \postcode{02130}, \country{Finland}}}
\affil[3]{\orgdiv{
Nano and Molecular Systems Research Unit},
\orgname{University of Oulu},
\orgaddress{\street{Pentti Kaiteran katu 1}, \city{Oulu}, \postcode{90570}, \country{Finland}}}

\affil[4]{\orgdiv{Department of Condensed Matter Physics and Materials Science}, \orgname{Tata Institute of Fundamental Research}, \orgaddress{Homi Bhabha Road}, \city{Mumbai}, \postcode{400005}, \country{India}}

\affil[5]{\orgname{VTT Technical Research Centre of Finland Ltd}, \orgaddress{\street{Tietotie 3}, \city{Espoo}, \postcode{02150}, \country{Finland}}}


\date{\today}

\abstract{
We theoretically introduce a quantum-circuit synthesis protocol to realize any $n$-qubit transformation $U\in U(2^n)$, up to a diagonal. The resulting circuit is directly native to and optimal in gate count and depth for systems where bitstring-conditioned rotations with tunable angle and axis, $C^{(n-1)}R(\theta,\phi)$, form the set of native operations. Subsequently, we suggest how this protocol can be implemented in a few-mode superconducting circuit, and experimentally demonstrate the method for the smallest nontrivial case, i.e., for two qubits. We report an average gate fidelity of ~97.6\% for Haar-random two-qubit unitaries up to phase shifts in the computational basis. Importantly, these results suggest that multimode devices, such as those in the multimon or multiunimon families may become competitive with qubit-array networks for general-unitary implementations for qubit counts as low as three.
}

\maketitle

\section*{Introduction}
\label{sec:introduction}

In the noisy intermediate-scale quantum (NISQ) computing era~\cite{Preskill2018quantumcomputingin}, errors on physical qubits play a dominating role in determining whether a quantum algorithm can outperform the best known classical solution. Hence, in recent years, significant efforts have been made to improve hardware performance through, for example, device design, fabrication methods, and new material platforms~\cite{tuokkola2025methods, bland2025millisecond,kono2024mechanically,zwanenburg2013silicon,noiri2022fastdot,xue2022quantum,sah2024decay}, optimized control techniques~\cite{wesdorp2026mitigating,hyyppafastdrag,AnCon2011, werninghaus2021leakage,DRAG,NAT,xue2022quantum,sah2024decay}, and novel qubit types~\cite{dualrail,hyyppaunimon,naturequasicharge,PRXBifluxon}. Moreover, advanced quantum-circuit synthesis and compilation methods have been developed in order to minimize the number of errors arising in the exact or approximate implementation of quantum algorithms, given specific hardware constraints~\cite{Mottonen2006,QSD,Krolsynthesis,kuo2021quantumarchitecturesearchdeep,Khatri2019quantumassisted,Wang_2026,furrutter2024quantum,amy2013meet,fosel2021quantum,Pozzirouting,Zhang_2022,sivarajah2021t,drăgoi2026efficientcircuittranspilationcommuting}.

While scalable quantum algorithms typically exploit structure and do not necessarily require the implementation of arbitrary $n$-qubit unitary matrices, small dense unitaries can arise as efficient subroutines within larger algorithms~\cite{Montanaro2016,nielsen00,Shor,skosana2021demonstration,Grover,Kitaev1995QuantumMA,gilyen2019quantum,isometries2016}. To this end, dense unitaries can serve as a useful worst-case benchmark, and a primitive for small quantum models. Consequently, the synthesis of quantum circuits implementing general unitaries is an active field of research~\cite{tucci1998rudimentary,CHEN2013853,krol2022efficient,fischer2023universal,brennenqudit}. However, owing to the exponential number of elementary gates required~
\cite{min_CNOT_count}, implementing arbitrary dense $n$-qubit unitaries poses a prohibitively difficult task beyond a few qubits with contemporary hardware performance.

Since $4^n-1$ free parameters must be controlled with signals delivered from the control electronics to the quantum system to implement a general transformation in $SU(2^n)$, at least $4^n-1$ unavoidable imperfections arise from control operations. Furthermore, interaction with the environment subjects the qubits to non-unitary errors that accumulate with the total circuit execution time. Execution time, in turn, is dependent on the number of consecutive control operations, or circuit depth. As the circuit depth grows, qubits suffer decoherence during more consecutive control operations, sometimes merely idling while other qubits are being controlled.

Simple parameter counting gives a hard limit on the minimum number of native gates required to implement a general $n$-qubit unitary, based on the available number of free parameters carried by each layer of a given native gate. In qubit-array networks where single qubits connect to some number of others via special couplers, only single-qubit gates and controlled-not (CNOT) gates or other pairwise-entangling operations are typically allowed. The number of such two-qubit gates in a quantum circuit is a particularly important metric in the NISQ era, since two-qubit gates are typically at least an order of magnitude more prone to errors than single-qubit gates~\cite{aseguinolaza2024error,IQMhf,isometries2016}. With the native operations limited to single-qubit gates and CNOTs, parameter counting gives a minimum of $\frac{1}{4}(4^n-3n-1)$ CNOTs to implement any element of $SU(2^n)$~\cite{min_CNOT_count}, given all-to-all connectivity. Assuming all $
\left \lfloor{\frac{n}{2}}\right \rfloor 
$ disjoint CNOTs are parallelized, the asymptotic depth scaling is $\Omega (4^n/n)$, though no such depth-optimal decomposition has been reported in the literature. Moreover, if the connectivity network is sparser than all-to-all, most pairwise operations need to be further decomposed to comply with the hardware. With realistic assumptions, some number of qubits idle during each CNOT, and the depth is further increased. 
In contrast, when more complex entangling operations are allowed, parallelization can become more efficient. In particular, with parallelized $(n-1)$-fold controlled one-parameter rotations $\mathrm{C}^{(n-1)}\mathrm{R}(\theta)$, up to $2^{n-1}$ independent parameters can be delivered in a depth-one layer, resulting in the depth scaling $\Omega (2^{n+1})$ for generic unitaries~\cite{CSD-original}. 

Recently, compact multimodal superconducting quantum devices such as the trimon~\cite{Trimon-original,trimon-second,multimon}, and the multiunimon~\cite{sasu-multiunimon} have emerged as compelling platforms for implementations of complex few-qubit subroutines that correspond to small, dense unitaries. Both devices employ three nonlinear electrical modes as qubits with strong longitudinal $ZZ$ coupling. The coupling is always on, and each qubit has four transition frequencies, one for each state of the other two qubits. Thus, the native operation is a doubly controlled rotation $\mathrm{CCR}(\theta,\phi)$, where both the rotation angle, $\theta$, and the azimuthal angle of the rotation axis, $\phi$, can be controlled by adjusting the drive parameters. Furthermore, controlled rotations conditioned on different control-register bitstrings can be parallelized by overlaying drive pulses at different carrier tone frequencies~\cite{Trimon-original,trimon-second,maurya2026universalhamiltoniancontrolplanar}. While $\mathrm{CCR}(\theta,\phi)$ fidelities exceeding 99.9~\% have been demonstrated on the trimon~\cite{maurya2026universalhamiltoniancontrolplanar}, to date no experimental demonstration of the multiunimon has been reported in the literature.

In this work, we introduce a modified version of the quantum-circuit synthesis algorithm in Ref.~\cite{CSD-original}, taking full advantage of the two available control parameters in a $\mathrm{CCR}(\theta,\phi)$ rotation. Under the assumption that all rotations conditioned on distinct control-register eigenstates can be executed in parallel and with independently tunable amplitudes and phases, we achieve a circuit depth of exactly $2^{n}-1$ to implement any element of $U(2^n)$, up to an unobservable global phase and $2^{n}-1$ phase factors which do not affect measurement probabilities in the computational basis and can hence be omitted, at least if the circuit is to be immediately followed by a measurement. If the circuit serves as a subroutine of a larger circuit, the final diagonal is not superfluous, but may be absorbed into later parts of the circuit, and such processes may be continued until measurements are carried out.
Given the native operation $\mathrm{C}^{(n-1)}\mathrm{R}(\theta,\phi)$, the depth
$2^n-1$ is optimal, as shown in the next section.
In addition, we include a rotation-angle optimization protocol to minimize gate errors in practical systems, as well as an alternative scheme that is likewise optimal and uses $2^{n+1}-2$ depth-one gates with each elementary operation implementing a $\pi/2$ rotation. Furthermore, we execute a preliminary experimental demonstration of the angle-optimized minimal-depth method on a trimon device.

\begin{figure*}[h!]
\begin{center}
\includegraphics[width=0.7\textwidth]{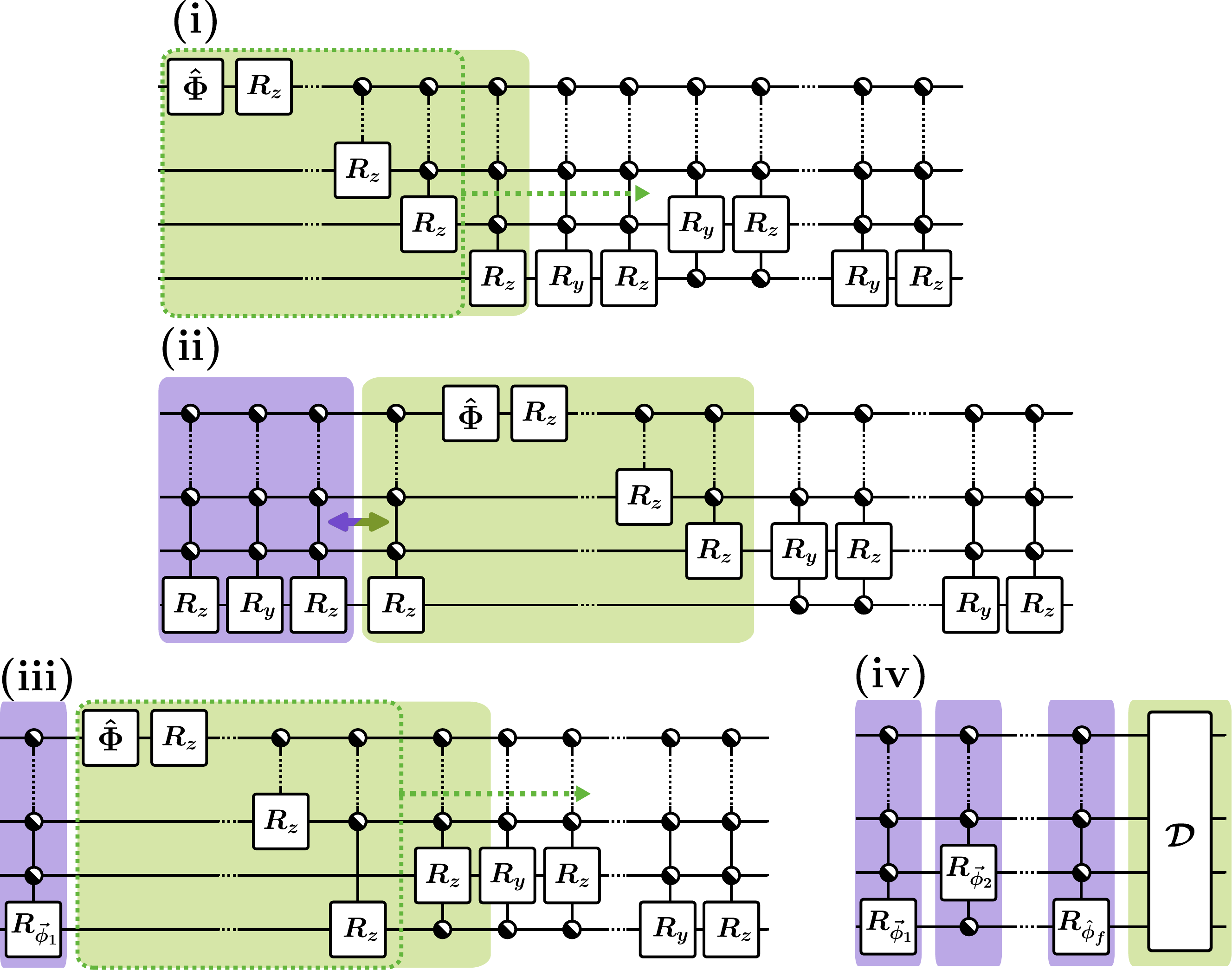}
 \end{center}
\caption{\textbf{Minimal-depth circuit synthesis based on the cosine-sine decomposition (CSD)}.
 \textbf{(i)}:
 The standard CSD-based circuit synthesis terminates with a diagonal $n$-qubit transformation shown as the green segment, followed by a sequence of $\UCRY\UCRZ$ pairs. Since phase shifts do not affect the probabilities of the computational states, they commute with control knobs, and the cascade inside the dashed square (the initial diagonal without the $(n-1)$-fold controlled UCR) commutes with the $\UCRZ\UCRY\UCRZ$ segment immediately following it. \textbf{(ii)}: The commuting part of the diagonal gate is moved to the other side of the $\UCRZ\UCRY\UCRZ$ segment, and the rightmost $\UCRZ$ is partitioned in two such that each rotation angle $\theta_i$ on the left side of the $\UCRY$ is matched with $-\theta_i$ on the right side. The remaining $\UCRZ$ is absorbed into the diagonal segment (green). \textbf{(iii)}: The resulting $\UCRZ\UCRY\UCRZ$ segment is a direct sum of rotations about axes in the $xy$-plane of the Bloch sphere. Here, $\vec{\phi}_1$ is an ($n-1$)-component vector specifying the azimuthal angle of the rotation axis for each basis state of the control register. Each of these rotations is a native operation in the scheme we consider, and can be parallelized with rotations conditioned on other eigenstates of the control register to form a depth-one gate. Next, the green cascade is restructured with the target qubit of the $(n-1)$-fold controlled $\UCRZ$ now matching the target qubit of the following $\UCRY$. \textbf{(iv)}: The commuting segment is moved across the next $\UCRZ\UCRY\UCRZ$ segment, the opposing $\UCRZ$s are anti-symmetrized as before, and the resulting sequence is merged into a single restricted uniformly controlled single-qubit gate (RUCG). The process is continued until the cascade reaches the end of the quantum circuit. Since the cascade, here represented as a single diagonal transformation, only affects phases, it does not change the measurement probabilities, and can be dropped, assuming the circuit is to be immediately followed by a measurement. Here $R_{\vec{\phi}_f}=R_{\vec{\phi}_{2^n-1}}$.\label{fig:decomp}}
\end{figure*}
\section*{Quantum-circuit synthesis protocols}
\subsection*{Minimal-depth circuit}
\label{sec:decomp}
The standard circuit synthesis protocol based on the cosine-sine decomposition (CSD)~~\cite{CSD-original,stewart1982computing} decomposes a general $n$-qubit unitary $U\in U(2^n)$ into a sequence of $2^{n}-1$ pairs of uniformly controlled rotations ($\mathrm{UCR}$s), with each pair consisting of one $\mathrm{UCR}$ about the $y$-axis ($\UCRY$), and one $\mathrm{UCR}$ about the $z$-axis of the Bloch sphere ($\UCRZ$). This sequence of pairs is immediately preceded or followed (depending on implementation) by a diagonal $n$-qubit transformation. Here, we present a simple formal protocol for transpiling the UCR circuit to a native gate set comprising $(n-1)$-fold controlled rotations $\mathrm{C}^{(n-1)}\mathrm{R}(\theta,\phi)$.

Our transpilation process is described in detail in Fig.~\ref{fig:decomp}. Starting with the decomposition obtained from the CSD such that the diagonal transformation lies at the beginning of the circuit, we push the diagonal past the $\UCRY\UCRZ$ pairs, leaving behind non-commuting parts, and conjugating the $\UCRY$ gates with such non-commuting parts. Each conjugation $\UCRZ\UCRY\UCRZ$ is then expressed as a uniformly controlled single-qubit gate (UCG)~\cite{UCGBergholm}. Here, the $\mathrm{UCG}$ is a direct sum of rotations about axes in the $xy$-plane. We coin these $\mathrm{UCG}$s restricted to rotations about axes in the $xy$-plane \textit{restricted uniformly controlled single-qubit gates} ($\mathrm{RUCG}$s). The elementary rotations making up an $\mathrm{RUCG}$ are indeed $\mathrm{C}^{(n-1)}\mathrm{R}(\theta,\phi)$. 

Furthermore, rotations conditioned on different eigenstates of the control register commute, meaning they can be parallelized, and the whole $\mathrm{RUCG}$ can, at least theoretically, be implemented as a depth-one gate. The general unitary is thus ultimately expressed as $2^{n}-1$ sequential $\mathrm{RUCG}$s, each consisting of $2^{n-1}$ parallelized native operations, with the whole sequence followed by a diagonal $n$-qubit unitary $\mathcal{D}(2^n)$. Assuming $U(2^n)$ is to be directly followed by a measurement in the computational basis, $\mathcal{D}(2^n)$ is irrelevant, as it does not affect projections onto eigenstates, and may be omitted, i.e. $U(2^n)\sim\mathcal{D}^{-1}(2^n)U(2^n)$. Thus, the effective final depth is 
$2^n-1$.

The operation count and depth in this construction are optimal with regard to the native operation $\mathrm{C}^{(n-1)}\mathrm{R}(\theta,\phi)$. We prove this in two steps. First, a depth-one layer can contain at most $2^{n-1}$ rotations. To see this, consider an $n$-dimensional hypercube whose vertices represent the computational basis states and whose edges connect states differing in a single bit. For example, for $n=2$, this graph is simply a square whose vertices are the four computational basis states and whose edges are the four single-bit transitions, of which at most two, namely a pair of opposite edges, are disjoint. Since the rotation axis lies in the $xy$-plane and all $n-1$ remaining
qubits act as controls, each $\mathrm{C}^{(n-1)}\mathrm{R}(\theta,\phi)$
corresponds to a transition along a single edge. Because rotations on edges sharing a vertex do not commute, the rotations in a single layer must occupy pairwise disjoint edges, of which there can be at most $2^n/2 = 2^{n-1}$. Hence, an RUCG is maximally parallelized. Second, at least $2^{n}-1$ consecutive RUCG layers are required. This is evident from simple parameter counting: a generic element of $U(2^n)$ has $4^n = 2^n(2^n-1)+2^n$ free parameters. Here, the first term is exactly the number of free parameters delivered by $2^n-1$ RUCGs with $2^n$ free parameters each, and the $+2^n$ term corresponds to the relative final phases and a global phase, all of which we omit. This parameter counting argument also shows that the number of elementary $\mathrm{C}^{(n-1)}\mathrm{R}(\theta,\phi)$ rotations is minimal.

\subsection*{Coterminal map}
\label{sec:coterm}
\label{optangles}
In a practical setting, small rotation angles are often preferable to larger ones since power dissipation and errors due to leakage out of the computational subspace typically increase with the drive pulse amplitude, and longer pulse durations lead to an increased effect of decoherence during a gate~\cite{krantz2019quantum,BlaisCQE}. Hence, we introduce a simple formal protocol for wrapping all angles in $(-2\pi,2\pi]$ into the principal interval $(-\pi,\pi]$, for implementations where this is not already done during the CSD step.

\begin{figure}[h!]
\begin{center}
\includegraphics[width=0.5\columnwidth]{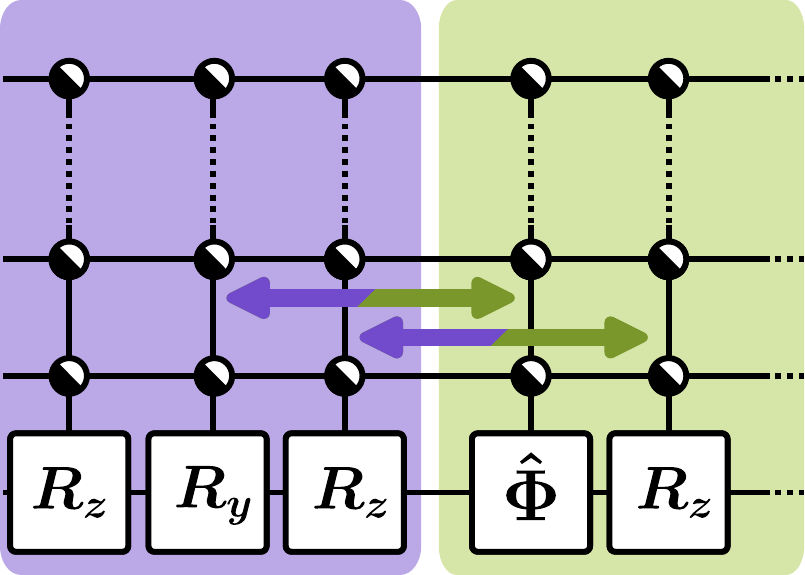}
\hfill
 \end{center}
\caption{\textbf{Mapping rotation angles to $(-\pi,\pi]$}. During step \textbf{(ii)} of Fig. \ref{fig:decomp}, every $\theta_i$ such that $|\theta_i|>\pi$ is replaced by its coterminal angle in $(-\pi,\pi]$. A phase gate conditioned on $n-1$ control qubits is inserted to flip the spinor phase where an angle was substituted for its coterminal. The phase gate is absorbed into the traveling diagonal gate. \vspace{0.5cm}\label{fig: coterm}
 }
\end{figure}

During the symmetrization step in Fig.~\ref{fig:decomp}, we map each rotation angle $\theta_i$ in the central $\UCRY$ to its principal coterminal angle in $(-\pi,\pi]$, and insert after it a diagonal sign matrix $\operatorname{diag}\!\{(-1)^{\ell_1},\ldots,(-1)^{\ell_n}\}$, where $l_i\in \{0,1\}$ are selected appropriately to correct the phase flips arising from the $2\pi$ angle changes, where necessary. The diagonal sign matrix commutes with the $\UCRZ$ on the right side, and is absorbed into the green segment in Fig.~\ref{fig:decomp} \textbf{(ii)}, which affects the azimuthal angles of the following $\mathrm{RUCG}$. This process illustrated in Fig. \ref{fig: coterm} is then repeated at each symmetrization step. 
\subsection*{Decomposition into $\pi/2$ rotations}
\label{sec: pi2}
\begin{figure}[h!]
\begin{center}
\includegraphics[width=1\linewidth]{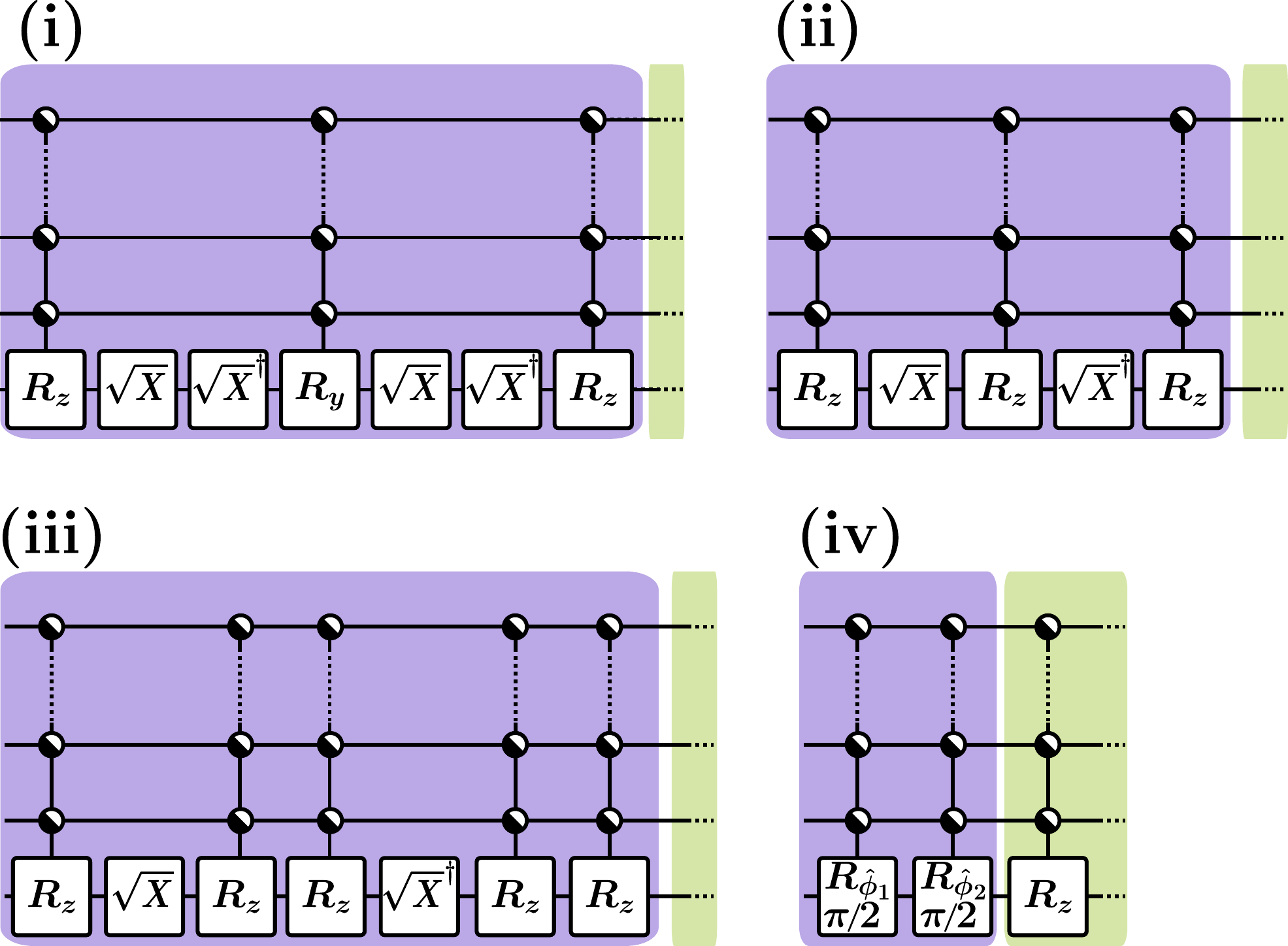}
\hfill
 \end{center}
\caption{\textbf{Decomposition in terms of $\pi/2$ rotations}. \textbf{(i)}: Identity operators $\sqrt{X}\sqrt{X}^\dag$ are inserted inside a $\UCRZ\UCRY\UCRZ$ segment. \textbf{(ii)}: Conjugation maps the rotation axis of the middle UCR from $y$ to $z$. \textbf{(iii)}: The $\UCRZ$ gates are partitioned into two such that the rotations on opposing sides of the single-qubit gates are inverse of each other. \textbf{(iv)}: Conjugation maps the rotation axis from $x$ to an axis in the $xy$-plane, specified by the azimuthal angle $\phi$. Here, the rotation angle is $\pi/2$ regardless of axis. The leftover $\UCRZ$ is absorbed into the traveling diagonal gate. Ultimately, the quantum circuit has $2^{n+1}-2$ $\mathrm{RUCG}$s, with every rotation angle exactly $\pi/2$.\label{fig: pi2}
 }

 \end{figure}

With a few additional steps visualized in Fig.~\ref{fig: pi2}, one can express the quantum circuit in terms of 
$\mathrm{RUCG}$s consisting of only $\pi/2$ rotations, i.e., direct sums of gates $\mathrm{C}^{(n-1)}\mathrm{R}(\pi/2,\phi)$. Since each  $\mathrm{RUCG}$  has only $2^{n-1}$ free parameters (one azimuthal angle for each eigenstate of the control register), the total operation count and circuit depth are doubled, but remain optimal with respect to the chosen elementary operations. This means that only one control operation per transition needs to be calibrated in order to implement any $n$-qubit transformation, assuming the azimuthal angle can be set with a calibration-free reference-frame update, as is often the case in quantum-gate implementations through resonant driving~\cite{EffZgate}. To map the whole circuit into $\pi/2$ rotations, identities $\sqrt{X}\sqrt{X}^\dag$ are inserted inside the $\UCRZ\UCRY\UCRZ$ segment during step \textbf{(ii)} of Fig.~\ref{fig:decomp} in order to conjugate the central $\UCRY$ into a $\UCRZ$, and to further conjugate the single-qubit gates into $\mathrm{RUCG}$s. 

\section*{Experimental demonstration}
\begin{figure}[!t]
\begin{center}
\includegraphics[width=0.7\linewidth]{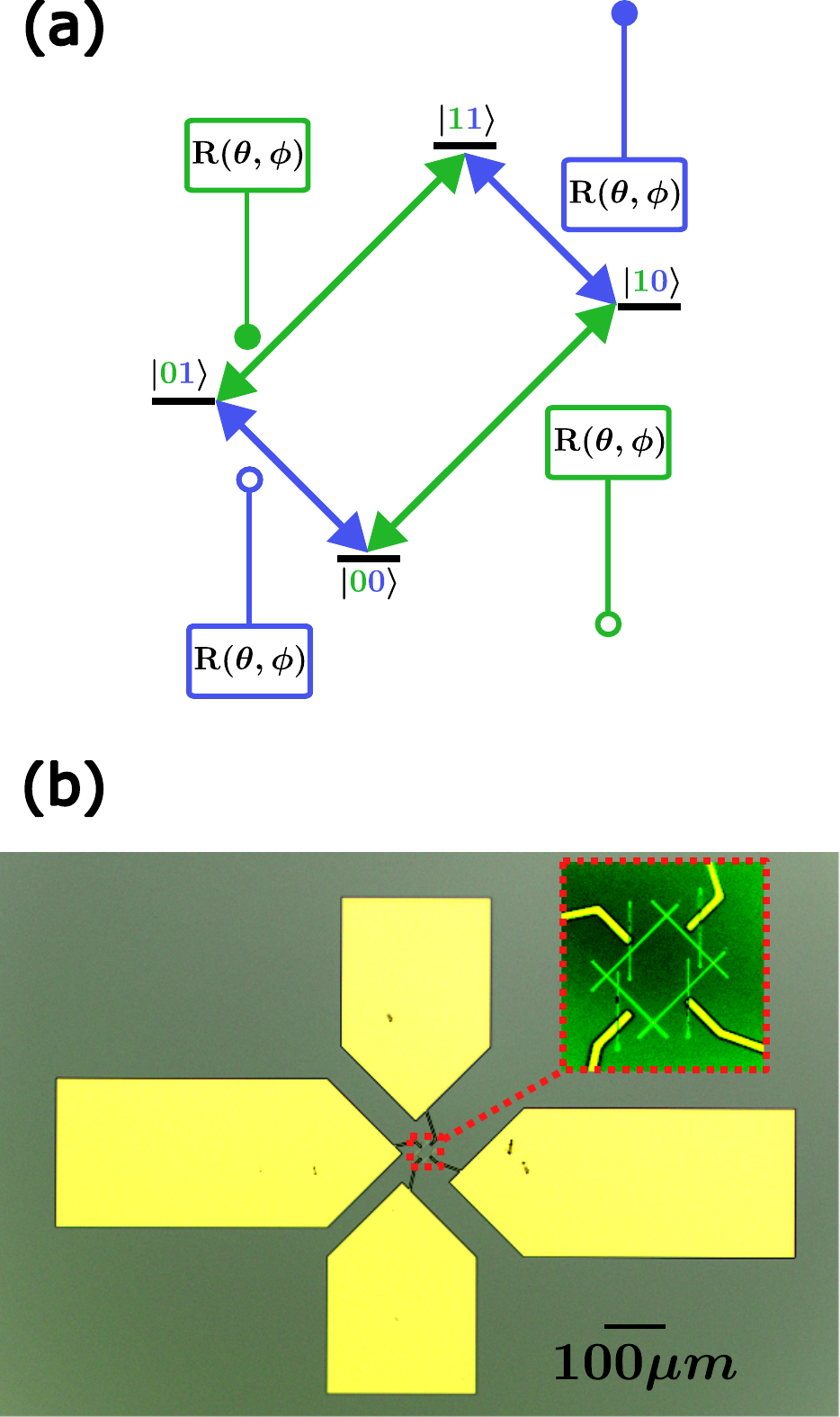}
\end{center}
  \caption{\textbf{Computational states and sample.} \textbf{(a)}: Energy levels and transitions in the computational space used in the experiments. The blue and green arrows correspond to transitions of qubits 1 and 2, respectively. Driving a transition implements a controlled rotation $\mathrm{CR}(\theta,\phi)$; since both edges corresponding to transitions of the same qubit are disjoint, they can be driven simultaneously to implement a $\mathrm{RUCG}(\theta_1,\theta_2,\phi_1,\phi_2)$.  
  \textbf{(b)}: Optical-microscope image of the device with enhanced contrast. A four-junction loop (inset) lies in the middle of the capacitor pads (yellow areas defined by pentagons).}
    \label{fig: el}
 \end{figure}
  \begin{figure*}[!t]
\begin{center}
\includegraphics[width=1\textwidth]{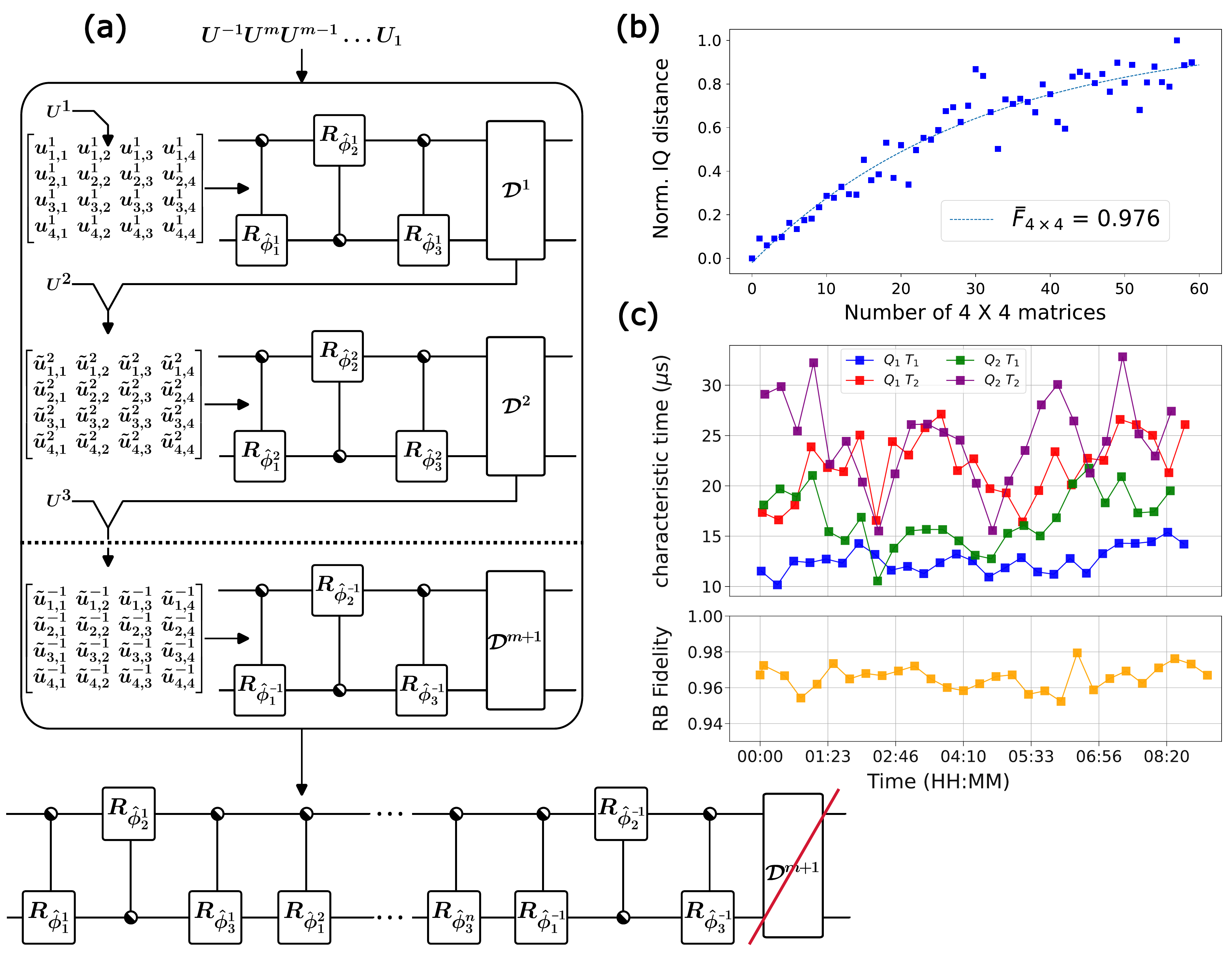}
\hfill
  \caption{\textbf{Randomized benchmarking protocol and results.} \textbf{(a)}: For each $m$, a sequence of $m$ Haar-random $4\times 4$ unitaries $U_i$ is decomposed into a quantum circuit, starting from $U_1$. The diagonal transformation at the end is absorbed in the next matrix. Finally, a matrix $U^{-1}$, the inverse of $U_mU_{m-1}...U_{1}$, is merged with the diagonal leftover from the preceding sequence, decomposed into a circuit, and the final diagonal is eliminated. Since each one of the $m$ matrices is decomposed into a circuit of depth three and each depth-one gate consists of two two-parameter control pulses $\mathrm{CR}(\theta,\phi)$, the total number of two-parameter elementary operations to implement $m$ $4\times 4$ unitaries is $6m$.  \textbf{(b)}: Normalized distance of the measured average signal of the two qubits in the in-phase--quadrature (IQ) space to the initial state $\ket{00}$ as a function of the number of $4\times 4$ matrices used in the RB sequence of panel (a). The exponential fit (dashed blue line) yields the depolarizing parameter associated with implementing a single general $4\times 4$ matrix. Each sequence of length $m$ is sampled independently and implemented $N_\mathrm{avg} = 10,000$ times. The corresponding fidelity is calculated according to Ref.~\cite{PhysRevLett.106.180504}. \textbf{(c)}: Mean randomized-benchmarking (RB) fidelity for a general random unitary transformation decomposed into a quantum circuit of depth three as a function of time (bottom) together with the energy relaxation time $T_1$ and the echo dephasing time $T_2$ between $\ket{01}$ and $\ket{00}$ ($Q_1$) or $\ket{10}$ and $\ket{00}$ ($Q_2$) (top).}
    \label{fig:fid}
    \end{center}
 \end{figure*}
We demonstrate the minimal-depth circuit with minimized rotation angles in a two-qubit space using a superconducting trimon~\cite{Trimon-original} device in a three-dimensional copper cavity. The trimon is a compact multiqubit circuit based on the transmon~\cite{transmon}, the most widely used type of superconducting qubit today. In a trimon, three spatially orthogonal dipole moments form three strongly ZZ-coupled qubits. Concretely, our device, shown in Fig.~\ref{fig: el} (b), consists of a loop of four Al-AlOx-Al tunnel junctions galvanically connected to Nb capacitor pads. Although the trimon ideally provides a three-qubit computational space, we find in our sample nearly overlapping transition frequencies which, given the short limiting coherence time and the resulting need for short, spectrally wide control pulses, make high-fidelity control involving the state $\ket{111}$ infeasible. We therefore carry out our demonstration in a two-qubit subspace, illustrated in Fig.~\ref{fig: el} (a). Mind that Fig.~\ref{fig: el} (a) is the $n=2$ physical realization of the hypercube-edge picture used in the quantum-circuit optimality argument. We number the qubits based on transition frequency, with qubit 1 the lower frequency qubit, and qubit 2 the higher frequency one. In bra-ket notation, we write the state of the most significant qubit on the left, i.e. $\ket{Q_2Q_1}$. The state of both qubits can be simultaneously determined with a measurement of the readout cavity, as described in Ref.~~\cite{Trimon-original}.

To determine the fidelity to which we can implement an arbitrary two-qubit transformation, we apply randomized benchmarking (RB)~\cite{RB-original,helsen2022general} of $4\times 4$ Haar-random unitary matrices, using the protocol shown in Fig.~\ref{fig:fid} (a). Each matrix is compiled into three $\mathrm{RUCG}$s, each of which is a direct sum of two $\mathrm{CR}(\theta,\phi)$ rotations. After decomposing a matrix, we merge the final diagonal with the next one, eventually pushing it to the very end of the whole sequence, where it is eliminated. 

We use 64 ns drive pulse envelopes for all transitions and superimpose the two drive pulses corresponding to a single $\mathrm{RUCG}$ in the decomposition. The rotation angles are controlled by independently adjusting the drive strengths. During a gate, we compensate for alternating-current (AC) Stark shifts and other drive-induced frequency shifts~\cite{ACstark,krantz2019quantum,walsh-hadamard} by dynamically updating the reference frame or drive frequency for each transition $\mathrm{t}$, based on frequency functions 
$f_\mathrm{t}(\vec{\theta})$, where $\vec{\theta}\in \mathbf{R}^4$ 
contains the rotation angles of all transitions during the $\mathrm{RUCG}$, half of which are zero. These functions are built in preliminary calibration before the fidelity measurement as described in the Methods section. Based on the RB fidelity time sequence data shown in Figs.~\ref{fig:fid} (c),
we report a best-decile fidelity of $97.6\pm 0.2~\%$ for a general $4\times 4$ unitary. Since each unitary is implemented as three $\mathrm{RUCG}$s, the average individual $\mathrm{RUCG}$ fidelity is roughly 
\begin{align}
\label{eq:CR_fid}
    \overline{F}_{\mathrm{RUCG}}\approx\sqrt[3]{0.976}\approx 99.2~\%.
\end{align}

\section*{Discussion}
In this work, we introduced a quantum-circuit synthesis protocol for systems where $(n-1)$-fold controlled rotations about axes in the $xy$-plane of the Bloch sphere form the set of native operations. Furthermore, we provided the first experimental demonstration of this protocol in a two-qubit subspace on a trimon device. Although high-fidelity two-qubit transformations have been demonstrated on superconducting qubit-array processors~\cite{Chen2025efficient}, systems of three qubits and beyond suffer such a significant disadvantage from the typically small number of different native entangling operations, that
corresponding multimode systems hold potential for fidelity improvements.

The $T_1$-type energy relaxation times in our device are relatively short, with the three-qubit state $\ket{111}$ surviving only a few microseconds on average, which, together with closely spaced transitions, prevented us from studying the three-qubit case. Nevertheless, one may use the presented results to derive estimates for larger systems.  Our two-qubit experiment implies a best-decile general three-qubit transformation ($8\times 8$ matrix) fidelity $\overline{F}_{8\times 8}\approx 93.3\%$, after removal of errors owing to finite $T_1$ and $T_2$. Alternatively, $\overline{F}_{8\times 8}\approx 91.6\%$ assuming $T_1=T_2=50$~{\textmu}s for all qubits. See Methods for a detailed description of the fidelity estimations. With an average $\mathrm{RUCG}$ fidelity of about 99.2~\%, the control-related error in our experiment is significant, which we attribute to the relatively large AC Stark shifts and the complex calibration and dynamic correction processes. Given longer coherence times, one may use longer control pulses and hence significantly reduce the control-related error, improving total fidelity. A purely coherence-limited fidelity for a three-qubit system with $T_1=T_2=50$~{\textmu}s and a depth-7 decomposition yields a fidelity limit of $\overline{F}_{8\times 8}\approx 97.2~\%$ for $100$~ns $\mathrm{RUCG}$s. On the other hand, a state-of-the art unconditional X gate fidelity $99.56~\%$ experimentally demonstrated on a trimon~\cite{maurya2026universalhamiltoniancontrolplanar}, gives $\overline{F}_{8\times 8}\approx 97.0~\%$, or $\overline{F}_{8\times 8}\approx 94.0~\%$ if the $\pi/2$-decomposition is used, though it is unlikely the best reported average unconditional X gate fidelity is directly applicable in the case of general RUCGs and initial states where complicated inter-transition AC Stark shift patterns become relevant.

For comparison, to implement a general $8\times 8$ matrix up to a diagonal, the best currently known synthesis method native to all-to-all coupled qubit arrays results in 18 two-qubit gates, and about 50 physical single-qubit $\sqrt{X}$-type gates~\cite{Krolsynthesis,Mottonen2006}. For simplicity, let us consider the total circuit fidelity in a somewhat optimistic scenario of 18 two-qubit gates implemented in a depth of 18 with one spectator qubit, and 50 physical single-qubit gates, with error-free $R_z$ single-qubit rotations, and no spectator errors during single-qubit gates. For superconducting qubits, a state-of-the-art single-qubit-gate fidelity of $99.99\%$, a two-qubit-gate fidelity of $99.9\%$~\cite{hfcz,IQMhf}, and one idle spectator qubit during each two-qubit gate obtaining a quarter of the infidelity of the participating qubits yield the estimate $\overline{F}_{8\times 8}\approx(0.999\times 0.99975)^{18}\times 0.9999^{50}\approx 97.3\%$. With 28 nearest-neighbor two-qubit gates~\cite{Krolsynthesis,Mottonen2006}, $\overline{F}_{8\times 8}\approx 96.1\%$. A more detailed explanation for the spectator error estimate is given in Methods.

For systems beyond three qubits, the difference in circuit depth between qubit-array networks and multimodal systems becomes very large. For example, at $n=4$, the method presented here yields a depth of 15, whereas the methods in~\cite{Krolsynthesis} yield 95 sequential CNOTs, assuming all-to-all connectivity. Three qubit systems therefore form the contemporary edge case where multimodal systems and qubit-array networks can implement general unitaries to comparable fidelity.
Thus, small multimode devices, especially the multiunimon~\cite{sasu-multiunimon} seem promising for demonstrating beyond state-of-the-art fidelity for general multiqubit gates when $n\geq 3$, which may improve the execution of compatible quantum algorithms.

\section*{Methods}
\subsection*{Compiling circuits}
The compilation of matrices into executable operations is implemented using a modified version of the Qcompiler program~\cite{CHEN2013853} (CSD), and a custom Python program (transpilation to $\mathrm{CR}(\theta,\phi)$ gates). The Python program is available at https://doi.org/10.5281/zenodo.23468224.

\subsection*{Fidelity estimates}
\label{sec:fidelity}
The average RB fidelity $\overline{F}_\text{RB}$ for our experiment is calculated from the RB decay parameter according to
Ref.~\cite{PhysRevLett.106.180504}.

The coherence-limited fidelity for a general $n$-qubit circuit is estimated as~\cite{NIELSEN2002249}
\begin{align}
\overline{F}_\text{coh}&=\frac{2^nF_{e,\text{coh}}+1}{2^n+1},\\
F_{e,\text{coh}} &= \prod_{i=1}^n F_{e,i},\\
F_{e,i}&=\frac{1+\ee^{-dt_\text{g}/T_{1,i}}+2\ee^{-dt_\text{g}/T_{2,i}}}{4},
\end{align}
where $t_\text{g}$ is the duration of a single native operation, and $d$ is the depth of the quantum circuit such that $dt_\text{g}$ is the total duration of the circuit. To estimate the control error for an $n$-qubit RUCG, we assume  the corresponding infidelity $1-\overline{F}_{\text{RUCG,ctrl}}$ doubles each time a new qubit is added, i.e., each time the number of control parameters doubles.

To estimate the spectator error during a two-qubit gate in a qubit-array system, we assume that control error and decoherent error contribute equally to the two-qubit gate error:
\begin{align}
1-\overline{F}_\mathrm{CNOT}\equiv r_\mathrm{CNOT}= r_\mathrm{decoh}+r_\mathrm{ctrl},\\ r_\mathrm{decoh}\approx r_\mathrm{ctrl},
\end{align}
and that the decoherent error during the gate is the sum of two equal decoherent errors on two qubits. Thus, the decoherent error on a single qubit during a gate, which we use as an estimate for the spectator error, is
\begin{align}
    r_\mathrm{spectator}\approx r_\mathrm{CNOT}/4.
\end{align}

\subsection*{Dynamic frequency shift correction}
\label{sec:sts_cal}
\begin{figure*}[!t]
     \begin{center}
 \includegraphics[width=0.9\linewidth]{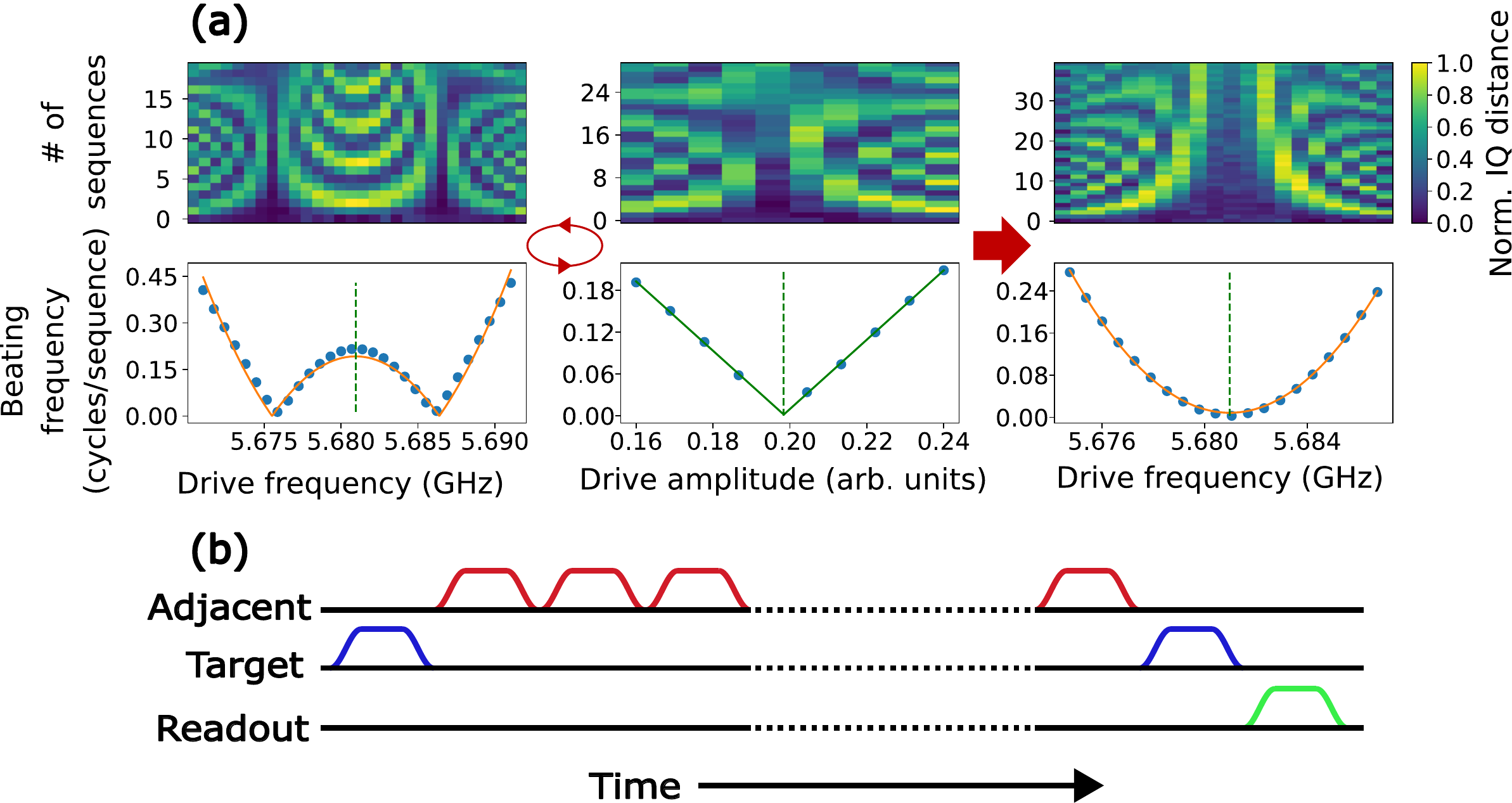}
     \hfill
     \end{center}
    \caption{\textbf{Frequency shift calibration.} \textbf{(a)}: Beating patterns and fits emerging in the process of calibrating the shift induced on the transition $\ket{00}\leftrightarrow\ket{10}$ while driving on resonance. The heatmap data corresponds to the normalized distance of the measurement signal of the two qubits in the in-phase–quadrature (IQ) space to the initial state $\ket{00}$. The first two panels show an initial frequency search and a following recalibration of the drive pulse amplitude, respectively. These two measurement are repeated multiple times until a clean beating pattern is acquired (last panel). \textbf{(b)}: Pulse sequence for frequency shift calibration between transitions with one common endpoint. The target qubit (blue) is initialized in an equal superposition of $\ket{0}$ and $\ket{1}$. While the target qubit idles, the adjacent transition (red) is rotated exactly by an integer multiple of the angle $4\pi$.  Finally, another $\pi/2$-pulse is applied to the target qubit, after which the system is measured with a readout pulse (green).\label{fig: sts}
    }
\end{figure*}
Our device suffers from relatively fast energy relaxation, with the average limiting coherence time about $7$~{\textmu}s in a two-qubit subspace. Hence, short gate durations are necessary in order to implement high-fidelity gates. This, in turn, leads to significant drive-induced frequency shifts such that a drive on a single transition shifts the frequency of every transition in the system. This is a known issue in multi-level superconducting quantum circuits~\cite{walsh-hadamard,maurya2026universalhamiltoniancontrolplanar}. 

In order to implement high-fidelity gates, the drive-induced frequency shifts must be calibrated to a very high precision, and corrected with dynamical updates. Furthermore, since we accept arbitrary rotation angles $\theta_i\in(-\pi,\pi]$, these corrections must be continuous functions of the magnitudes of the rotation angles, i.e. they depend only on $\{|\theta_1|,|\theta_2|,|\theta_3|,|\theta_4|\}$, two of which are zero during any given RUCG. To this end, we use an iterative calibration protocol to find the shifts for all drive-target pairs $\text{d}, \mathrm{t}\in{\{1,2,3,4\}}$, and a relevant collection of angles $\theta_\text{d}\in[0,\pi]$ as illustrated in Fig. \ref{fig: sts}. We then fit functions $f_\mathrm{t}(\theta_\text{d})= a\theta_\text{d}^2+f_\mathrm{t}(0)$, where $f_\mathrm{t}(0)$ is the target transition idle frequency, and $a\in\mathbb{R}$. Finally, we combine the results assuming the total shift during a multiplexed drive field is the sum of shifts induced by individual drives: $f_\mathrm{t}(\theta_1,\theta_2,\theta_3,\theta_4)=\left[\sum_{i=1}^4f_\mathrm{t}(\theta_i)\right]-3f_{\mathrm{t}}(0)$. 
The functions are cross-verified by checking that energy is conserved along different loops: for any set of rotation angles $\{\theta_i\}_{i=1,2,3,4}$, the total frequency difference in a closed loop (e.g. $\ket{00}\rightarrow\ket{01}\rightarrow\ket{11}\rightarrow\ket{10}\rightarrow\ket{00}$) must equal 0.

We distinguish three categories of calibration measurements, on the basis of the drive-target pair:
\begin{enumerate}
    \item When the drive and target correspond to the same transition, we perform the calibration by applying $k\in\{0,1,2,\ldots\}$ subsequences of $r\in\{2,3,...\}$ pulses, where an individual pulse is a first estimate to implement a gate $X^\frac{2}{r}$ on resonance. $k$ and drive frequency are varied to generate a beating fringe pattern, as shown in the top half of the first panel of Fig.~\ref{fig: sts} (a). Beating frequencies are calculated, and a periodically mirrored parabola shown in the bottom half of the first panel of Fig.~\ref{fig: sts} (a) is fitted. The vertex corresponds to resonance. Next, the drive frequency is fixed to the found resonance, while the pulse amplitude is varied. The fringe at $f_\text{beating}=0$ shown in the second panel of Fig.~\ref{fig: sts} (a) corresponds to a true rotation angle of $\theta=2\pi/r$. Since changing the pulse amplitude affects the frequency shift, the process is iterated from the beginning until a clean fringe is obtained. The entire process is then repeated for different sequence lengths and rotation angles defined by $r$; finally, a linear function is fitted between $(2/r)^2$ and the obtained frequencies. The protocol also generates precise instrument drive amplitude settings for desired rotation angles.
    \item When the drive and target correspond to transitions with disjoint endpoints, e.g. $\ket{00}\leftrightarrow\ket{01}$ and $\ket{10}\leftrightarrow\ket{11}$, the shift induced on the target transition can be calibrated with a simple Ramsey-type phase accumulation measurement where a continuous drive tone is applied on the other transition while the target idles. The measurement is repeated for multiple continuous drive amplitudes to obtain the target frequency as a linear function of the interleaved drive power.
    \item When the drive and target correspond to transitions with one common endpoint, for example $\ket{00}\leftrightarrow\ket{01}$ and $\ket{00}\leftrightarrow\ket{10}$,
    we conduct a Ramsey type phase accumulation measurement, where the other transition is rotated by an integer multiple of the angle $4\pi$ while the target idles. The exact pulse sequence is illustrated in Fig.~\ref{fig: sts} (b). Before the actual calibration, the drive pulse on the adjacent transition is carefully calibrated to exactly implement a gate $X^\frac{4}{r}$, $r\in\{4,5,\ldots\}$ on the adjacent transition. The actual frequency shift calibration is started by preparing the target qubit in an equal superposition of $\ket{0}$ and $\ket{1}$ with a $\pi/2$-pulse. $k\times r$, where $k\in \{0,1,\ldots\}$, pulses are applied to the adjacent transition to implement a rotation by the angle $k\times4\pi$, an integer multiple of $4\pi$ instead of $2\pi$ so as not to flip the spinor phase. Finally, a second $\pi/2$ pulse is applied to the target qubit and the system is measured. $k$ and the target idle frequency are varied to generate a fringe beating pattern, and the shifted qubit frequency is extracted. The experiment is then repeated for multiple values of $r$, and a linear function is fitted between the rotation angles $(4/r)^2$ and the found frequencies.
\end{enumerate}

Our frequency shift calibration process also generates precise instrument drive amplitude settings for desired rotation angles. In order to extrapolate this to a continuous $|\theta|\in [0,\pi]$, we fit a third degree heuristic polynomial approximation $A_\mathrm{t}(\theta)\approx b |\theta|+c |\theta|^2+d|\theta|^3$, where $b,c,d \in\mathbb{R}$, to the acquired data points. 

While the $\mathrm{RUCG}$ decomposition is exponentially shallower in $n$ compared to any decomposition native to qubit-array networks, the number of required frequency shift calibration measurements grows exponentially with the number of qubits. Currently, we are able to build each frequency function $f_\mathrm{t}$ in an hour of calibration. Since there are four transitions $\mathrm{t}$, the total time taken is approximately four hours. With three qubits, the number of functions $f_\mathrm{t}$ and the number of parameters in each of them increase from four to 12, meaning the total required duration is about 36 hours. Fortunately, opting for the $\pi/2$-decomposition and/or emerging fast calibration techniques~\cite{marciniak2026calibration,FastBayesiancal,RestlessTuneup2017} may significantly decrease the calibration overhead.

\subsection*{Sample fabrication}
The sample is fabricated in a process similar to that described in Ref.~\cite{tuokkola2025methods}. First, a 200-nm layer of Nb is sputtered on top of a 675-$\mathrm{\mu}$m-thick Si subtsrate and patterned using optical lithography and CF$_4$-plasma etching to define the capacitor pads. Then, we remove the oxide layers from the Nb and Si surfaces by immersing the sample in a 0.5\% hydrofluoric acid solution for 10 minutes, followed by a rinse in deionized water (DIW). The sample is directly transferred to a spincoater from the DIW and coated with 500 nm of copolymer resist consisting of methyl methacrylate–methacrylic acid having 11\% solids in ethyl lactate and 400 nm of poly(methyl methacrylate) (PMMA) 950K A4 resist (4\% solids in anisole; nominal molar mass approximately 950,000 g mol$^{-1}$). The resist stack is patterned using an electron-beam writer with 100-kV acceleration voltage, 300-$\mathrm{\mu}$m aperture and 0.5-nA beam current, applying a dose of 1000 $\mathrm{\mu C / cm^2}$ to create the line-like shapes and 400 $\mathrm{\mu C / cm^2}$ to create undercuts at the line ends. The resist mask is then developed in a 1:3 solution of methyl isobutyl ketone and isopropyl alcohol (IPA) for 5 minutes, rinsed in pure IPA for 1 min and dried with nitrogen. 

The Manhattan-style junctions are deposited in an electron-beam evaporator after overnight pumping of the chambers to a pressure below $7\times 10^{-8}$ mbar and ozone cleaning at 10 mbar for 1 minute. The first electrode and the insulating layer are created by evaporating 40 nm of Al at $45^{\circ}$ planetary angle, and oxidizing in-situ at 0.13 mbar for 5 minutes. For the second electrode, we evaporate 30 nm of Al at both $-45^{\circ}$ and $135^{\circ}$ planetary angles. To ensure clean contact to the Nb capacitor pads, we use Ar milling prior to the final deposition step of the connecting leads, followed by the evaporation of 30, 60, and 60 nm of Al at planetary angles $180^{\circ}$, $0^{\circ}$, and $180^{\circ}$, respectively. Each Al layer is deposited at $45^{\circ}$ tilt angle and a rate of 0.2 nm/s after Ti gettering and pumping the chamber pressure to or below $5.1\times10^{-8}$ mbar. Finally, the sample is oxidized at 20 mbar for 10 minutes to create a clean protective layer of aluminum oxide on top of the junctions.

After the junction deposition, the sample is spincoated with protective photoresist and pre-diced to $2\times 5$~mm$^2$ chips with 450-$\mathrm{\mu m}$ cutting depth. Finally, the sample is immersed in $80~^{\circ}$C Remover PG for 3 hours, and sonicated with medium power in fresh Remover PG, acetone, and IPA for 3 minutes each, followed by nitrogen drying and immediate transfer to vapor prime oven, where a protective monolayer of hexamethyldisilazane is applied on the sample surface with effective annealing at $150~^{\circ}$C for 20 minutes. The junction resistances are measured at room temperature with a probe station, and the sample is manually cleaved into individual chips.

\subsection*{Experimental setup and device parameters}
\begin{figure}[h!]
     \begin{center}
\includegraphics[width=1\linewidth]{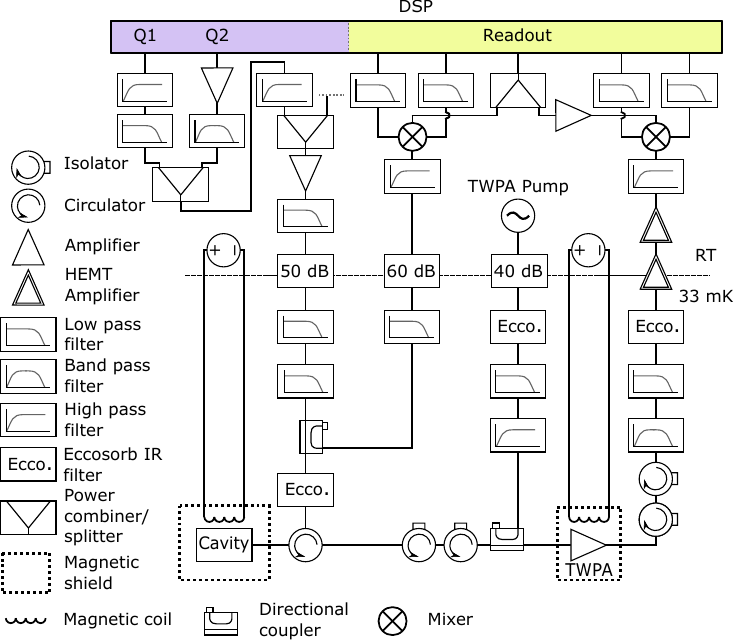}
     \hfill
    \caption{\textbf{Experimental setup}. TWPA denotes traveling-wave parametric amplifier; HEMT denotes high-electron-mobility transistor; IR denotes infrared. The room temperature (RT) and cryogenic electronics are separated by the horizontal dashed line. The readout and drive lines are separated before the mixing-chamber stage in order to attenuate frequencies close to that of the readout cavity resonance significantly more than the frequencies inside the drive band. Some of the components of the full three-qubit setup are omitted from the schematic as indicated by the dashed line before one of the power combiners in the RT part of the setup. Signal generation and data acquisition are handled by a digital signal processing module (DSP).\label{fig: setup}}
    \end{center}
\end{figure}

The experiments are conducted in a cryogenic dilution refrigerator at a base temperature of about 33~mK. The setup shown in Fig.~\ref{fig: setup} was initially built for experiments involving all three qubit modes, and left as is for the two-qubit experiments to be able to benefit from the already existing calibrations. The sample lies in a copper cavity placed inside a magnetic shield. The frequencies and anharmonicities of the qubits together with the cavity frequency are shown in table~\ref{table: freqs}.

\begin{table}[h!]
\caption{Frequencies and anharmonicities $\alpha$ of the qubit transitions and resonator cavity used in the experiment. N/A denotes not applicable.\label{table: freqs}}
\begin{tabular}{|c|c|c|}
\hline
Transition & Frequency (GHz) & $-\alpha$ (MHz)\\
\hline
$00\leftrightarrow01$ & 5.117& 128\\
    \hline
    $10\leftrightarrow11$& 4.863&\\
    \hline
    $00\leftrightarrow10$&5.935 &116\\
    \hline
    $01\leftrightarrow11$& 5.681&\\
    \hline
    Cavity&8.239 &N/A\\
    \hline
\end{tabular}

\end{table}
\section*{Data availability}
The data that support the findings of this study
are available at https://doi.org/10.5281/zenodo.23468224.
\section*{Author contributions}
H.S. and M.M. conceived the circuit synthesis protocols and the experiments. H.S. conducted electromagnetic simulations and designed the device based on the simulation data, and theoretical results from S.T. H.K. fabricated the device. H.S conducted the experiments with help and feedback from S.K, V.V., and R.V. H.S. analyzed the results. E.F. contributed to the initial characterization experiments and the experimental setup. M.T. contributed to the fabrication and design processes. M.C. fabricated an initial device. H.S. developed the circuit synthesis theory. H.S., H.K., and M.M. wrote the manuscript with comments from all the authors.
\section*{Acknowledgments}
We thank the KAUTE foundation and The Magnus Ehrnrooth foundation for their support through Ph.D. grants awarded to the first author. We acknowledge financial support from the Research Council of Finland through the Academy Professor's project Autonomous Quantum Machines (Grant No. $369679$), from the European Research Council under the Advanced Grant ConceptQ (No. $101053801$), from Jane and Aatos Erkko Foundation through the SystemQ project, from Novo Nordisk Foundation through the project Sustainable Quantum Computers (No. NNF$25$OC$0103322$), and from Horizon Europe programme HORIZON-CL4-2022-QUANTUM-01-SGA via the project 101113946 OpenSuperQPlus100. Furthermore, we are thankful to Dr. Miika Rasola, Tuomas Uusnäkki, Aashish Sah, Ashish Panigrahi, and Dr. Thomas Loke for their assistance and insightful discussions. 
\section*{Competing interests}
M.M. declares that he is a Co-Founder and Share-
holder of IQM Finland Oy and of QMill Oy. S.K. declares that he is a Co-Founder and Shareholder of Coherence Quantum Oy.
All other authors
declare no competing interests.
\bibliography{refs}
\end{document}